\documentclass[conference, letterpaper]{IEEEtran}
\IEEEoverridecommandlockouts
\usepackage{pdfpages}
\usepackage{balance}
\usepackage{gensymb}
\usepackage{cite}
\usepackage{amsmath,amssymb,amsfonts}
\usepackage{algorithmic}
\usepackage{graphicx}
\usepackage{textcomp}
\usepackage{xcolor}
\def\BibTeX{{\rm B\kern-.05em{\sc i\kern-.025em b}\kern-.08em
    T\kern-.1667em\lower.7ex\hbox{E}\kern-.125emX}}
\begin{document}

\title{Closed-Loop Ramp-Comparison Current Regulator for an Induction Machine with a PWM Voltage-Source Inverter\\}

\author{\IEEEauthorblockN{Aidar Zhetessov}
\IEEEauthorblockA{\text{Dept. of Electrical and Computer Engineering} \\
\text{1415 Engineering Drive}\\
Madison, WI 53706}
}

\maketitle

\begin{abstract}
This paper addresses the closed-loop ramp-comparison current regulation in an induction machine fed by a pulse width modulated voltage source inverter. The regulator is implemented in a synchronous frame, serving as a foundation for an overarching vector control of the induction machine. First, the effect of PI regulator gains on the controller performance is analyzed both theoretically and numerically using the developed Simulink model of the system. Next, the paper deals with high-speed and/or low-voltage operating conditions of the machine, introducing the concept of overmodulation and analyzing its impact on the regulator performance. Obtained simulation results coincide with model-based theoretical predictions and literature findings. Finally, the work proposes an outlook for the high-speed system enhancements in terms of power electronics topology, control and modulation.
\end{abstract}

\begin{IEEEkeywords}
induction machine, PWM, VSI, current regulator, overmodulation, synergetic control
\end{IEEEkeywords}

\section{Introduction}
Induction Machines (IM) are an integral part of a very diverse set of applications, benefiting our society on a daily basis. Traditional IM application areas include fans, pumps, compressors, conveyors etc. The list of emerging applications can be filled with wind turbine generators \cite{Abad} and electric vehicle motors \cite{Pellegrino}. All those applications (especially emerging ones) require a high-performance torque/speed regulation of the IM.

To address the regulation necessity, the high-performance IM control has been the subject of extensive research for several decades. In all these efforts, an established regulation standard has emerged both in terms of hardware and control. From hardware perspective, a Voltage-Source Inverter (VSI) power electronics topology is regarded as the standard solution for IM drive \cite{Novotny}. From control perspective, an Indirect Field Oriented Control (IFOC) regulator architecture is known as the standard \cite{Novotny}. As an IM drive solution, VSI with IFOC provide a high-performance machine speed/torque regulation. Moreover, given high-bandwidth (idealized) regulation, IFOC decouples qd-axes in synchronous frame, effectively deactivates rotor dynamics and ultimately allows tackling an inherently nonlinear IM model analytically \cite{Novotny}. The keyword here, though, is the \textit{idealized} regulation.

In its core, IFOC is a set of cascaded (predominantly PI) regulators operating in a synchronous frame and controlling the IM speed $\rightarrow$ torque/rotor magnetization $\rightarrow$ stator currents in cascaded manner respectively. The innermost current regulator terminates by a synchronous-to-stationary frame transformation and a subsequent modulator - a block that implements various modulation schemes to generate the required voltages at the stator terminals using the DC-link voltage of the VSI. These generated stator voltages are the ultimate control handles applied to the IM. The idealized regulation is achieved when the outer speed controller has a high bandwidth. This, in turn, poses strict requirements on the innermost current regulator and modulator. Therefore a high-bandwidth IM stator current regulation is crucial for a high-performance IM control within IFOC context.

Numerous works have addressed both current regulator design and modulation technique selection in the past. For instance, regarding controller, reference \cite{Novotny} mentions hysteresis-based current control, PI feedback regulators in stationary and synchronous frames, PI regulator model-based enhancements like feedforward compensation and even the family of predictive controllers. As of the modulators, various solutions are discussed in \cite{Holmes}, \cite{Hava}. This work addresses the simplest realization: two PI current regulators in a synchronous frame followed by a PWM modulator. Having grasped a firm understanding of the IM current regulation through the realization at hand, potential limitations could be identified and further modifications/improvements could be added. In fact, this paper already identifies some limitations of the simplest realization through simulations and proposes enhancements in terms of control/modulation and even power electronics topology.

The paper is organized as follows. Section \ref{Sec:Setup} describes the system setup that was used for analysis, both theoretical equations and Matlab/Simulink model. Section \ref{Sec:Regulator} addresses the PI current regulator design, expected and simulated performance. Section \ref{Sec:Overmodulation} deals with high-speed and/or low DC voltage operating conditions. Last but not least, Section \ref{Sec:Conclusion} summarizes the work and lists the potential improvements for high-speed operation.

\section{Setup Description}\label{Sec:Setup}
Matlab/Simulink provides a comprehensive environment for the closed-loop analysis of the induction machine - voltage source inverter (IM-VSI) system under consideration, thus it was used for the numerical simulations. The specifications of the IM of interest are summarized in Table \ref{t:params}.

\begin{table}[tbp]
	\caption{Induction Machine Parameters}
	\begin{center}
		\begin{tabular}{|c|c|c|}
			\hline
			\textbf{Machine}&\multicolumn{2}{|c|}{\textbf{Specifications}} \\
			\cline{2-3} 
			\textbf{Parameter}&\textbf{\textit{Value}}&\textbf{\textit{Description}}\\
			\hline
			$V_{ll,rms}$&460 V&Line-to-line rms voltage\\
			\hline
			$P$&20 hp&Rated machine power\\
			\hline
			$f$&60 Hz&Excitation frequency\\
			\hline
			$Pol$&4&Number of poles\\
			\hline
			$r_s$&0.355 $\Omega$&Stator resistance\\
			\hline
			$r_r$&0.355 $\Omega$&Rotor resistance\\
			\hline
			$X_{ls}$&1.42 $\Omega$&Stator leakage reactance\\
			\hline
			$X_{lr}$&1.42 $\Omega$&Rotor leakage reactance\\
			\hline
			$X_{m}$&34.1 $\Omega$&Magnetizing reactance\\
			\hline
			$M$&1.4 sec&Mechanical time constant\\
			\hline
			\textbf{Base}&\multicolumn{2}{|c|}{\textbf{Base Values}} \\
			\cline{2-3} 
			\textbf{Parameter}&\textbf{\textit{Value}}&\textbf{\textit{Description}}\\
			\hline
			$V_{B}$&376 V&Base voltage (peak)\\
			\hline
			$I_{B}$&26.5 A&Base current (peak)\\
			\hline
			$Z_{B}$&14.20 $\Omega$&Base impedance\\
			\hline
			$T_{B}$&79.16 N-m&Base torque\\
			\hline
			$\omega_{B}$&$2\pi f$ rad/s&Base angular speed\\
			\hline
			\textbf{Rated}&\multicolumn{2}{|c|}{\textbf{Rated Operation}} \\
			\cline{2-3} 
			\textbf{Parameter}&\textbf{\textit{Value}}&\textbf{\textit{Description}}\\
			\hline
			$s_{R}$&0.03135&Rated slip\\
			\hline
			$I_{sR}$&1.244 pu&Rated stator current (peak)\\
			\hline
			$cos(\theta_R)$&0.861&Rated power factor\\
			\hline
		\end{tabular}
		\label{t:params}
	\end{center}
\end{table}

\begin{figure*}[htbp]
	\centerline{\includegraphics[width=0.95\textwidth]{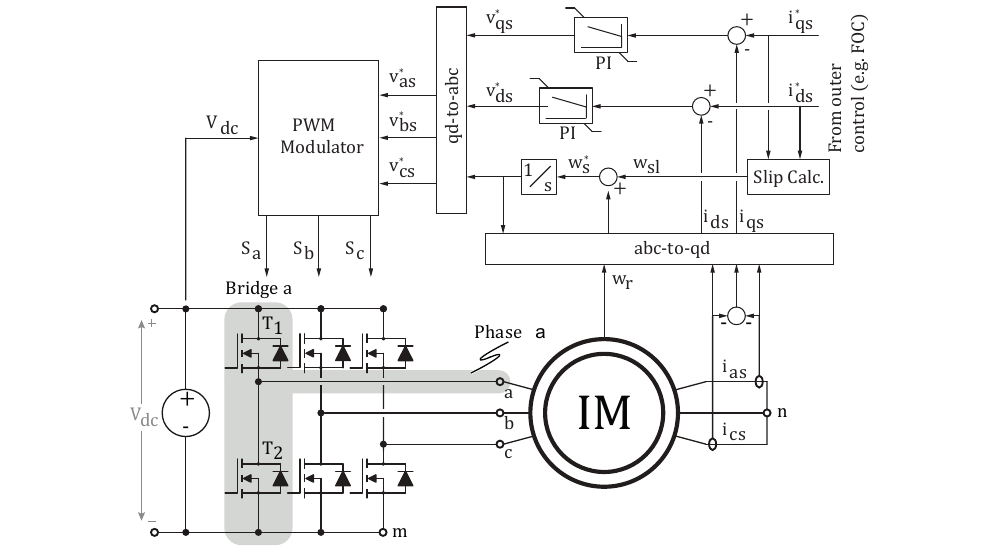}}
	\caption{High-Level Closed-Loop IM-VSI Block Diagram}
	\label{fig:model}
\end{figure*}

Using the specifications from Table \ref{t:params}, a closed-loop simulation was developed. A high-level block diagram of the developed model can be observed in Fig. \ref{fig:model}. The model comprises the IM, PWM inverter (PWM VSI) and the closed-loop ramp-comparison regulator structure. The IM is actuated by the VSI in stationary \textit{abc} frame. Stator current measurements, used for feedback, are also captured in \textit{abc} frame. Regarding the regulator, it was shown in \cite{Novotny},\cite{Rowan} that AC current control shows superior performance if implemented in synchronous \textit{qd}-frame rather than in stationary \textit{abc}-frame. Indeed, closed-loop PI regulators have unity gain and no phase shift at DC, rather than AC reference quantities. Moreover, with an advent of digital signal processing capabilities, changing reference frame does not pose a significant challenge. Finally, regulating all three currents in \textit{abc}-frame could result in undesired controller interactions, since there are only two linearly independent currents in the floating-neutral IM configuration \cite{Novotny}. With these being said, the current regulation is implemented in a synchronous \textit{qd}-frame.

The following subsections address the individual blocks of Fig. \ref{fig:model} in more detail.

\subsection{Induction Machine}
The developed IM model is based on the differential equations from \cite{Novotny}. Those equations, including auxiliary relations, are summarized in \eqref{eq:IMmodel}-\eqref{eq:IMaux}, where $\Psi_{qs},\Psi_{ds},\Psi_{qr},\Psi_{dr}$ states correspond to flux voltages, $\omega_r$ state corresponds to rotor speed, and $T_L$ corresponds to the load torque. Five differential equations (4 electrical, 1 mechanical) capture the dynamics of AC machines, including the nonlinear effects of state multiplications. Moreover, they provide an opportunity to capture the time-varying parameter effects in numeric simulations e.g. magnetizing inductance saturation $L_m = f(i_m)$ etc. For now, though, parameter variations are not captured in the model.

\begin{equation}
\begin{split}
&V_{qs} = r_s\Big(\frac{\Psi_{qs}-\Psi_{mq}}{X_{ls}}\Big) + \frac{d}{\omega_Bdt} \Psi_{qs} + \frac{\omega}{\omega_B}\Psi_{ds},
\\
&V_{ds} = r_s\Big(\frac{\Psi_{ds}-\Psi_{md}}{X_{ls}}\Big) + \frac{d}{\omega_Bdt} \Psi_{ds} - \frac{\omega}{\omega_B}\Psi_{qs},
\\
&V_{qr} = r_r\Big(\frac{\Psi_{qr}-\Psi_{mq}}{X_{lr}}\Big) + \frac{d}{\omega_Bdt} \Psi_{qr} + \frac{(\omega-\omega_r)}{\omega_B}\Psi_{dr},
\\
&V_{dr} = r_r\Big(\frac{\Psi_{dr}-\Psi_{md}}{X_{lr}}\Big) + \frac{d}{\omega_Bdt} \Psi_{dr} - \frac{(\omega-\omega_r)}{\omega_B}\Psi_{qr},
\\
&\frac{M}{\omega_B}\frac{d}{dt}\omega_r = \frac{(T_e-T_L)}{T_B}.
\end{split}
\label{eq:IMmodel}
\end{equation}

\begin{equation}
\begin{split}
&\frac{1}{X_m^*} = \frac{1}{X_m} + \frac{1}{X_{ls}} + \frac{1}{X_{lr}},
\\
&\Psi_{mq} = \frac{X_m^*}{X_{ls}}\Psi_{qs} + \frac{X_m^*}{X_{lr}}\Psi_{qr},
\\
&\Psi_{md} = \frac{X_m^*}{X_{ls}}\Psi_{ds} + \frac{X_m^*}{X_{lr}}\Psi_{dr},
\\
&i_{qs} = \big(\Psi_{qs}-\Psi_{mq}\big)/X_{ls},
\\
&i_{ds} = \big(\Psi_{ds}-\Psi_{md}\big)/X_{ls},
\\
&T_e = \frac{3}{2}\frac{Pol}{2}\frac{1}{\omega_B}\Big(\Psi_{ds}i_{qs}-\Psi_{qs}i_{ds}\Big).
\end{split}
\label{eq:IMaux}
\end{equation}

\begin{figure}[htbp]
	\centerline{\includegraphics[width=0.5\textwidth]{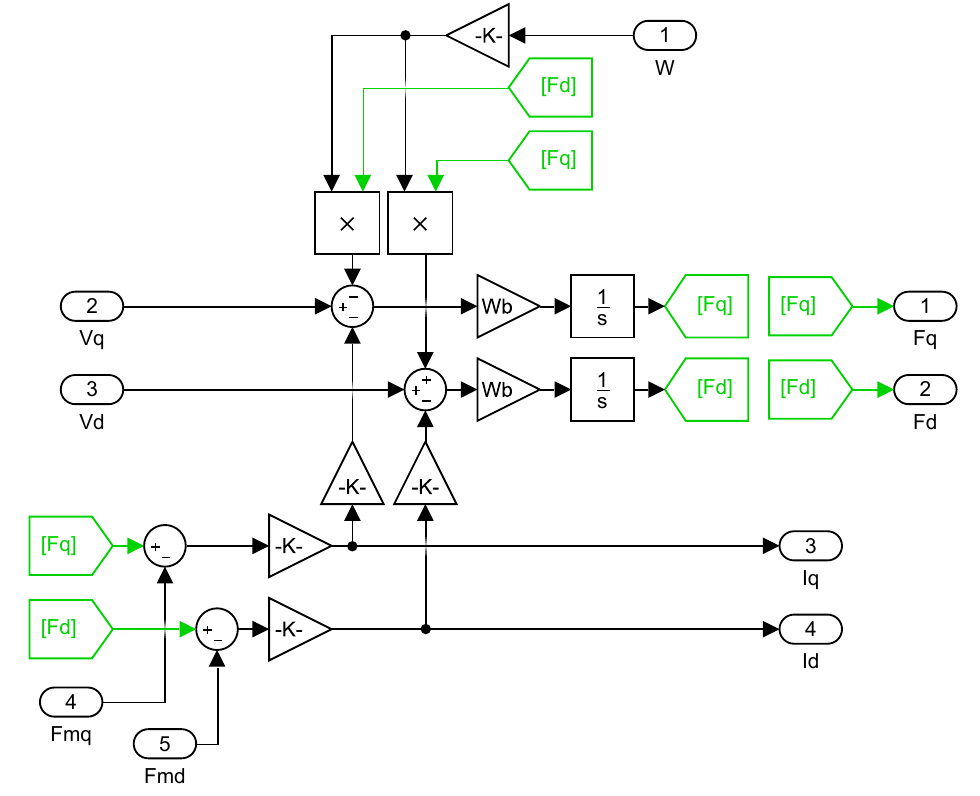}}
	\caption{Simulink Realization of Stator Voltage Equations}
	\label{fig:VqdsSimulink}
\end{figure}

An example realization of the first two stator equations of \eqref{eq:IMmodel} in Simulink are depicted in Fig. \ref{fig:VqdsSimulink}. Subsequent rotor electrical and mechanical equations are realized similarly. Note that the presented Simulink realization is in per unit (p.u.) notation, such that one has a convenient analysis/comparison tool for different induction machines.

\subsection{Voltage Source Inverter}
Within the developed Simulink model the VSI is realized as a two-step look up table. First, given the switching commands from the PWM modulator, the VSI maps the DC-link voltage $V_{dc}$ to the stator phase-to-negative voltages $V_{am},V_{bm},V_{cm}$. This relation can be expressed as follows in \eqref{eq:VSImap1}:

\begin{equation}
\text{for } i \in [a,b,c],V_{im} = 
	\begin{cases}
	V{dc},& \text{if } S_i=1\\
	0,    & \text{if } S_i=0
	\end{cases}
\label{eq:VSImap1}
\end{equation}

Second, phase-to-negative voltages $V_{am},V_{bm},V_{cm}$ can be mapped to the stator phase voltages $V_{an},V_{bn},V_{cn}$ and the neutral voltage $V_{nm}$ as follows in \eqref{eq:VSImap2}:

\begin{equation} 
\begin{split}
V_{nm} &= \frac{V_{am}+V_{bm}+V_{cm}}{3},\\
V_{an} &= V_{am} - V_{nm} = V_{as}^*,\\
V_{bn} &= V_{bm} - V_{nm} = V_{bs}^*,\\
V_{cn} &= V_{cm} - V_{nm} = V_{cs}^*.
\end{split}
\label{eq:VSImap2}
\end{equation}

\subsection{Current Regulator}
\begin{figure}[htbp]
	\centerline{\includegraphics[width=0.5\textwidth]{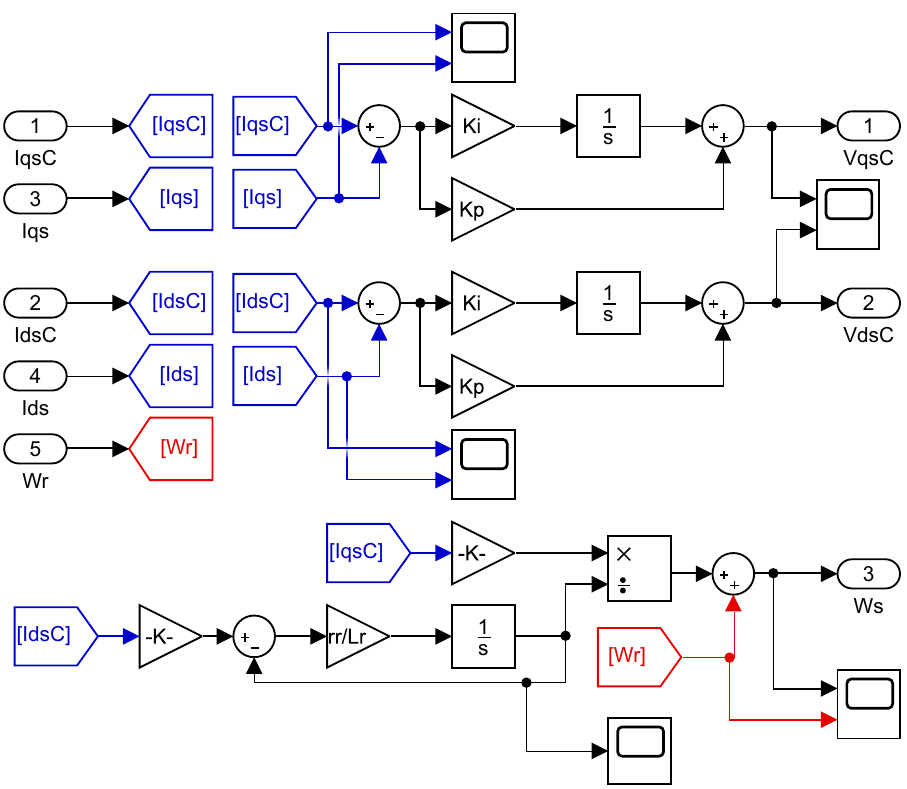}}
	\caption{Simulink Realization of Current Regulator in \textit{qd}-frame}
	\label{fig:IqdsSimulink}
\end{figure}

As Fig. \ref{fig:model} suggests, current regulator model is realized in a synchronous \textit{qd}-frame and comprises three major parts: q-component stator current controller, d-component stator current controller and a synchronous speed/angle calculator. Simulink realization of the current regulator model is shown in Fig. \ref{fig:IqdsSimulink}. From the perspective of control theory, qd-component current regulators fall into the series compensation category are responsible for closed-loop current control. Provided the errors of the stator current qd-components, the regulators output the stator voltage qd-component references for a subsequent modulator. No further enhancements, such as model-based feedforward and/or saturation/anti-windup are applied at this point.

The synchronous speed/angle calculator determines the \textit{abc/qd} transformation angle and, consequently, the IM excitation frequency. With this synchronous speed/angle calculator the rotor speed variations are accounted in control and the reference frame-rotor flux alignment is preserved. In essence, it allows applying the established Indirect Field Oriented Control (IFOC) on top of the presented current regulator in case rotor speed, torque and/or magnetization needs to be controlled \cite{Novotny}. As in IFOC, the equations of the slip calculator block in Fig. \ref{fig:model} (also captured in the bottom section of Fig. \ref{fig:IqdsSimulink}) are summarized in \eqref{eq:SlipCalc}:

\begin{equation} 
\begin{split}
L_m i_{ds}^{e*} &= \hat{\lambda}_{r} + \tau_r\frac{d\hat{\lambda}_{r}}{dt},\\
\omega_{sl} &= \frac{r_r}{L_r} \frac{L_m}{\hat{\lambda}_{r}} i_{qs}^{e*},\\
\omega_{s} &= \omega_{sl} + \omega_{r}.
\end{split}
\label{eq:SlipCalc}
\end{equation}

\subsection{PWM Modulator}
\begin{figure}[htbp]
	\centerline{\includegraphics[width=0.5\textwidth]{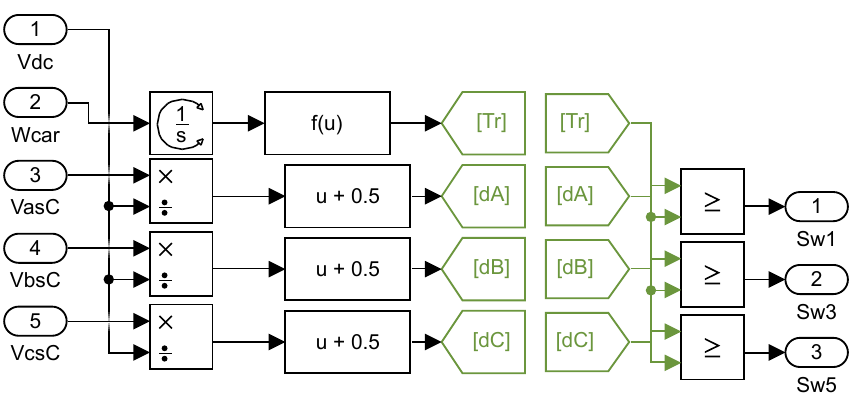}}
	\caption{Simulink Realization of PWM Modulator in \textit{abc}-frame}
	\label{fig:PWMSimulink}
\end{figure}

PWM Modulator block (Fig. \ref{fig:PWMSimulink}) works in \textit{abc}-frame and generates the switching signals for the VSI look up table. To do so, first it derives three duty cycles for three phases independently by dividing the regulator-generated voltage references to the DC-link voltage and adding a constant of $1/2$ to the result \eqref{eq:DutyCycles}: 

\begin{equation}
\text{for } i \in [a,b,c],d_{i} = \frac{V_{is}^*}{V_{dc}} + \frac{1}{2}
\label{eq:DutyCycles}
\end{equation}

It can be shown that by setting this addition value to a constant $1/2$, the average of the neutral voltage $V_{nm}$ is controlled to $V_{dc}/2$, implying a constant common-mode voltage of the set $V_{am},V_{bm},V_{cm}$. Within VSI context, the modulation strategy that results in a constant common-mode voltage of $V_{nm} = V_{dc}/2$ is referred to as a standard PWM. Of course, it is possible to shape the addition value in a non-constant fashion. This opens a whole field of various modulation techniques such as Discontinuous PWM (DPWM) and Space Vector Modulation (SVM), which are the subjects of aforementioned works \cite{Holmes}, \cite{Hava}. The underlying concept of duty cycles, however, still can be related to all of those. 

Next, the duty cycles are fed to the comparators. A saw-tooth carrier waveform is compared to the duty cycles to produce the discrete switching signals of 0 and 1. For the sake of simulation speed the carrier frequency was selected to be $f_{sw} = 100f$. Thanks to the fact that the carrier frequency is kept constant, the average switching frequency and the resulting harmonic content of the IM voltages and currents almost always (excluding overmodulation conditions) remain in the vicinity of the switching frequency and its integer harmonics \cite{Novotny}. Moreover, this ramp-comparison method excitation production results in a convenient sampling of the electrical measurements without the necessity of further post-processing \cite{Rowan}, \cite{Yu}.

\section{Current Regulator Design and Performance} \label{Sec:Regulator}
Having addressed the analytical model and simulation setup of the IM-VSI system, this section deals with PI current regulator design and its performance evaluation.

\subsection{Regulator Design}
\begin{figure}[tbp]
	\centerline{\includegraphics[width=0.5\textwidth]{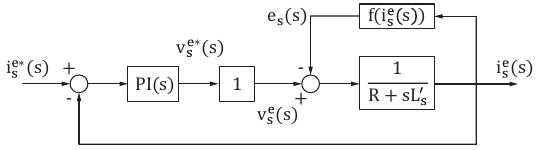}}
	\caption{Simplified block diagram of series-compensated current-regulated IM}
	\label{fig:RegulatedIMdiagram}
\end{figure}

Regarding the regulator design, in the developed setup with a series compensation-type PI regulators in \textit{qd}-frame, the task simplifies to proportional and integral gain tuning. This gain tuning, in turn, requires either some knowledge of the plant model in \textit{qd}-frame (frequency-domain analysis, transfer functions, Bode plots \cite{Novotny}) or some classical gain tuning method along with numerous simulations and experience (Ziegler-Nichols method \cite{Shin},\cite{Ogata}). An IM-VSI plant model, in form of a transfer function in \textit{qd}-frame, is derived in \cite{Novotny}, allowing for a straightforward controller design using an established frequency-domain toolbox. A simplified block diagram of the series-compensated current-regulated IM is shown in Fig. \ref{fig:RegulatedIMdiagram}. This block diagram serves both for q- and d-axes.

Here the VSI is modeled as a unity transfer function, while the IM is an $RL$ low-pass filter with a current-dependent feedback term, regarded as a disturbance \cite{Novotny},\cite{Shin}. From \cite{Shin} one knows that for IM the transfer function parameters are the following \eqref{eq:IMtransFunc}:

\begin{equation}
\begin{split}
R &= r_s + r_r \Big(\frac{L_m}{L_r}\Big)^2,\\
L_s' &= \sigma L_s = L_s - \frac{L_m^2}{L_r}.
\end{split}
\label{eq:IMtransFunc}
\end{equation}

Having IM transfer function (TF) defined, the open-loop TF (loop gain $T_{ol}(s) = i_s^e(s)/i_s^{e*}(s)$) can be derived as \eqref{eq:LoopGain}:

\begin{equation}
\begin{split}
T_{ol}(s) &= \frac{i_s^e(s)}{i_s^{e*}(s)} = \Big(K_P + \frac{K_I}{s}\Big)\Big(\frac{1}{R + sL_s'}\Big)\\
&= \frac{K_P}{s}\Big(s+\frac{K_I}{K_P}\Big)\Big(\frac{1/L_s'}{R/L_s' + s}\Big)
\end{split}
\label{eq:LoopGain}
\end{equation}

As suggested in \cite{Shin}, for pole-zero cancellation one should select the gain ratio of $R/L_s'$, which results in the following open-loop and closed-loop TFs \eqref{eq:LoopGainFinal}:

\begin{equation}
\begin{split}
\frac{K_I}{K_P} &= \frac{R}{L_s'},\\
T_{ol}(s) &= \frac{K_P}{sL_s'},\\
T_{cl}(s) &= \frac{T_{ol}(s)}{1+T_{ol}(s)} = \frac{K_P}{K_P + sL_s'}.
\end{split}
\label{eq:LoopGainFinal}
\end{equation}

By tuning $K_P = \omega_c L_s'$, one selects the cross-over frequency $\omega_c$ of the closed-loop TF, thus defining the bandwidth of the controller. A rule of thumb for $\omega_c$ selection is one-tenth of the switching frequency ($2\pi f_{sw}/10$). On one hand, the rule leaves a decade of margin to the switching frequency such that the regulator is not too aggressive, while one the other hand the rule ensures high controller bandwidth and disturbance rejection at low frequencies.

\begin{figure}[tbp]
	\centerline{\includegraphics[width=0.5\textwidth]{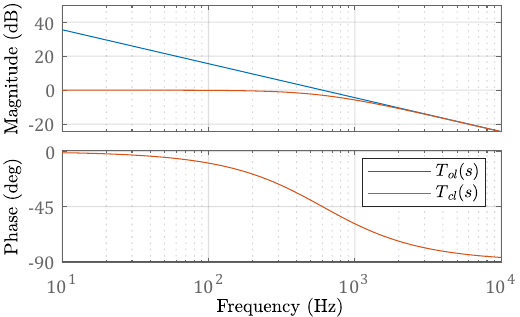}}
	\caption{Open- and closed-loop transfer function Bode plots}
	\label{fig:BodeTcl}
\end{figure}

For the selected $\omega_c = 2\pi f_{sw}/10$, open- and closed-loop TF Bode plots are depicted in Fig. \ref{fig:BodeTcl}. The corresponding PI gains are $K_P = \omega_c L_s'$ and $K_I = \omega_c R$. From Fig. \ref{fig:BodeTcl} it can be observed that the closed-loop TF has a cross-over frequency (bandwidth) at $f_c = f_{sw}/10 = 600$ Hz. As of the phase margin, it is equal to $\phi=90\degree>0\degree$, implying stability due to phase margin criterion. Indeed, the phase margin is defined as $\phi = 180\degree + \angle T_{ol}(j\omega_c)$. Given the fact that $T_{ol}(j\omega_c)=-90\degree$, the phase margin is well above zero.

\subsection{Regulator Performance}
To evaluate the effects of the designed regulator, and particularly its bandwidth, on the regulation performance, a load torque step was simulated. For the scenario at hand the DC voltage was selected to be $V_{dc} = 2.5V_B$ rather than a PWM limit of $2V_B$ to be on safe side in terms of undesired overmodulation (that will be addressed in the next section). The step occurs after three out of six fundamental periods and drops from $1$ p.u. to $0.5$ p.u. In this simulated scenario the electrical torque is controlled to follow the load step immediately. This is done through the step in q-component stator current reference $i_{qs}^{e*}$ from its rated value of $1.184$ p.u. to about half of it. Note that the rated value of $i_{qs}^{e*}$ is slightly below the $I_{sR}$ value of $1.244$ p.u. (Table \ref{t:params}) due to existence of the stator current d-component $i_{ds}^{e*}$ in the selected \textit{qd}-frame aligned with rotor flux linkage (IFOC context). The aforementioned $i_{ds}^{e*}$ does not change after load step to preserve the rotor magnetization.

\begin{figure}[htbp]
	\centerline{\includegraphics[width=0.5\textwidth]{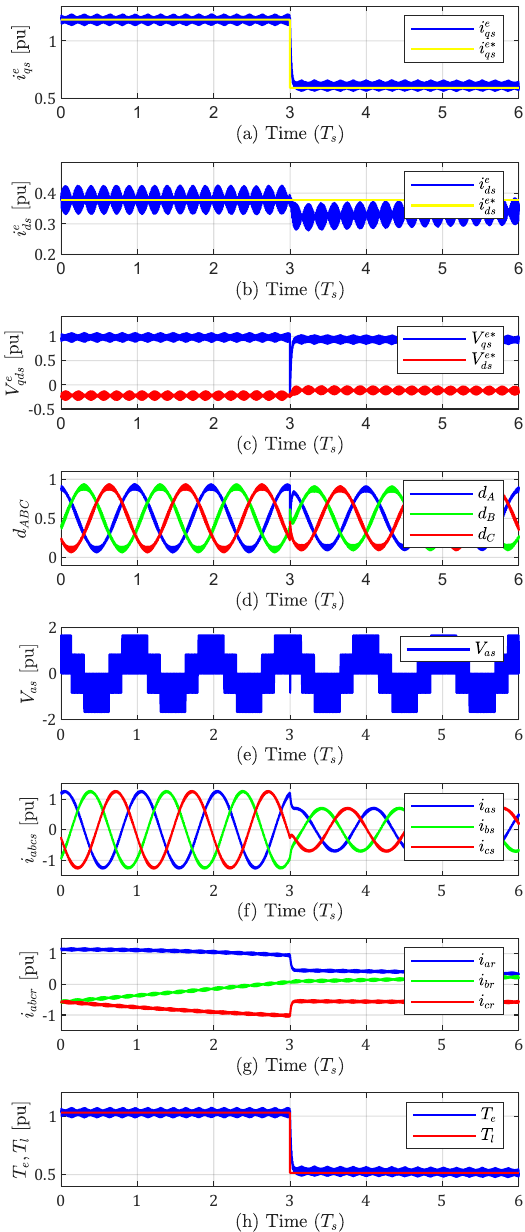}}
	\caption{Load and torque step simulation waveforms for $V_{dc} = 2.5V_B$. qd-component stator currents and their references $i_{qds}^e,i_{qds}^{e*}$ (a-b), qd-component stator voltage references $V_{qds}^{e*}$ (c), \textit{abc}-frame duty cycles (d), phase \textit{a} stator voltage (e), \textit{abc}-frame stator and rotor currents (f-g), load and electrical torques (h).}
	\label{fig:Performance1}
\end{figure}

From Fig. \ref{fig:Performance1} it can be seen that current regulators are fast in responding to the reference step, which translates to almost instantaneous electrical torque control. Both qd-component voltages and duty cycles appear ripply, meaning that controller bandwidth is high. Indeed, ripply duty cycles imply that some sidebands of the switching frequency lie below the regulator bandwidth such that it tries to suppress those switching ripple sidebands to a certain extent. Nevertheless, duty cycle ripples are tolerable in the scenario at hand (always less than $1$, $V_{dc}>2V_B$) so there is no need to sacrifice the controller bandwidth.

Numerous simulations with lower bandwidth values $\omega_c$ indicate that current regulators react in a slower fashion, gradually landing form one steady-state ($t = 0$) to the other ($t = 6T_s$). Voltage references in \textit{qd}-frame and duty cycles appear less and less ripply, ultimately converging to the ideal sine waveforms. Obviously, high-frequency ripples in qd-current components and electrical torque still persist due to switching operation of the VSI, which is now completely uncontrolled. Here it is worth mentioning that, in addition to high-frequency ripple, low-frequency harmonic content (small waves) in qd-currents and qd-voltages is presumably due to $V_{am},V_{bm},V_{cm}$ common-mode voltage $V_{nm}$ containing harmonics besides PWM-prescribed DC value of $V_{dc}/2$.

On the other hand, higher bandwidth values $\omega_c$ result in more and more aggressive regulation at the price of higher and more non-sinusoidal duty cycles even during the steady-state operation of the first three periods. As the bandwidth approaches the switching frequency, generated voltage references and duty cycles resemble sine waves less and less ultimately leading to instability in the vicinity of $f_{sw}$. This is expected, since at $f_{sw}$ the averaged VSI model of unity transfer function (Fig. \ref{fig:RegulatedIMdiagram}) is no longer valid and the phase margin argument is no longer true. 

Overall, it was shown that the designed current regulator achieves a high-performance control as predicted by the TF-based analysis (Fig. \ref{fig:BodeTcl}) and by the findings in \cite{Shin}.

\section{High-Speed and/or Low Voltage Operation} \label{Sec:Overmodulation}
The underlying assumption of the last section was that the VSI DC voltage is large enough so that the voltage references from the designed current regulator could be generated by the inverter. In particular, for the PWM modulator at hand the DC voltage was selected to be $V_{dc} = 2.5V_B > 2V_B$. This section, in turn, addresses the conditions when the VSI is not able to produce the commanded AC voltages immediately because the DC voltage is too low and/or the regulator voltage command is too high. Similar situation was briefly mentioned in the discussion of an aggressive regulator tuning.

The stator equations of \eqref{eq:IMmodel} can be used to derive the stator voltage in steady-state, synchronous frame and complex notation \eqref{eq:Vqds}:

\begin{equation}
\underline{V}_{qds}^e = r_s \underline{i}_{qds}^e + j\frac{\omega}{\omega_B}\underline{\Psi}_{qds}^e
\label{eq:Vqds}
\end{equation}

From \eqref{eq:Vqds} one can observe that $\underline{V}_{qds}^e$ is a vector summation of the resistive voltage drop and the speed voltage term. On the other hand, from \eqref{eq:VSImap1} it can be derived that for the PWM linear modulation interval \cite{Novotny} the length of the stator voltage space vector is limited by the DC voltage as $|\underline{V}_{qds}^e| \leq V_{dc}/2$. Combining the last two equations leads to the inequality condition for remaining in the linear PWM modulation region \eqref{eq:VqdsCondition}:

\begin{equation}
\frac{V_{dc}}{2} \geq |\underline{V}_{qds}^e| = |r_s \underline{i}_{qds}^e + j\frac{\omega}{\omega_B}\underline{\Psi}_{qds}^e|
\label{eq:VqdsCondition}
\end{equation} 

Compliance with the condition \eqref{eq:VqdsCondition} means that the PWM VSI can operate in the linear region of modulation index \cite{Novotny} and that the DC voltage is large enough to generate the required stator voltage space vector. Conversely, for a given stator current space vector (rotor magnetization and torque) the condition can be violated when the synchronous speed $\omega$ is too high or, equivalently, when the DC voltage $V_{dc}$ is too low. Violation of the condition implies that PWM modulator enters the nonlinear interval, which manifests in duty cycles exceeding the carrier limits of $0$ and $1$ ($d_{ABC} > 1, d_{ABC} < 0$). This is referred to as \textit{overmodulation}. In overmodulation conditions modulator starts dropping pulses. As the required AC voltage commands grow, the VSI AC voltage waveforms converge to a six-step operation - a theoretical maximum limit of producible fundamental AC voltage amplitude. 

Generally, it is possible to extend the linear modulation interval of PWM by a proper shaping of common-mode voltage $V_{nm}$. Moreover, there are other achievable objectives, such as reduced VSI switching loss, number of switching events etc. All those are addressed by more advanced modulation techniques such as Space Vector Modulation (SVM), PWM with third-harmonic injection, Discontinuous PWM (DPWM) etc.

\begin{figure}[htbp]
	\centerline{\includegraphics[width=0.5\textwidth]{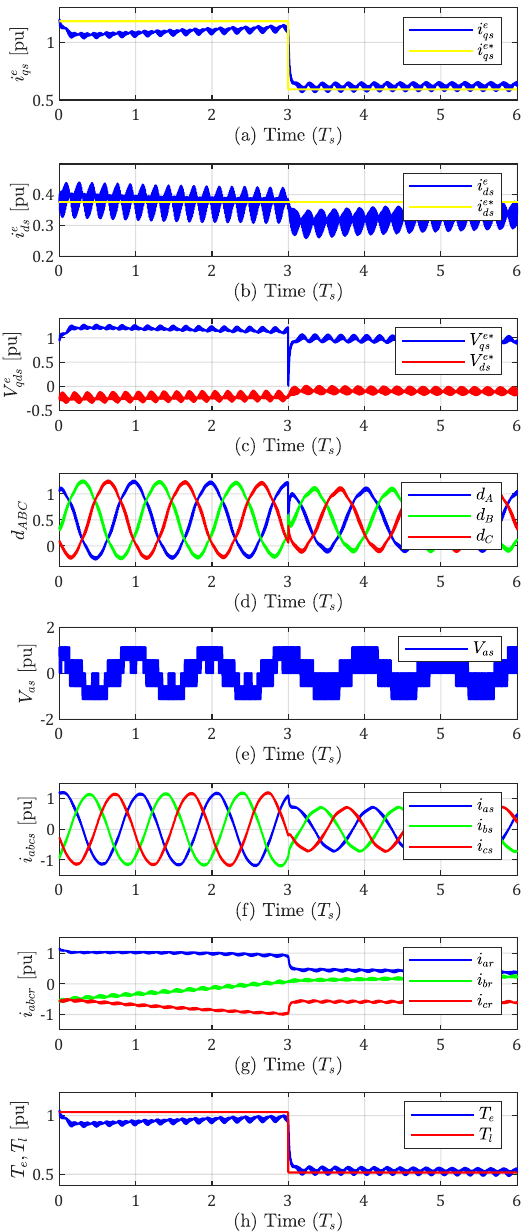}}
	\caption{Load and torque step simulation waveforms for $V_{dc} = 1.7V_B$. qd-component stator currents and their references $i_{qds}^e,i_{qds}^{e*}$ (a-b), qd-component stator voltage references $V_{qds}^{e*}$ (c), \textit{abc}-frame duty cycles (d), phase \textit{a} stator voltage (e), \textit{abc}-frame stator and rotor currents (f-g), load and electrical torques (h).}
	\label{fig:Performance2}
\end{figure}

Strictly speaking, overmodulation is another example of nonlinearity that affects the unity TF model of the VSI in Fig. \ref{fig:RegulatedIMdiagram}. Nevertheless, provided the regulator does not push the VSI too far into the nonlinear region, the qd-current references can still be tracked as it can be seen in Fig. \ref{fig:Performance2}.

Fig. \ref{fig:Performance2} simulates the same scenario as Fig. \ref{fig:Performance1}, but now with DC voltage of only $V_{dc} = 1.7V_B < 2V_B$. This inevitably pushes the VSI into overmodulation as the $V_{as}$ waveform (Fig. \ref{fig:Performance2}e) suggests. From Fig. \ref{fig:Performance2}a one can see that initially the VSI is not able to provide the requested excitation due to low DC voltage and, consequently, stator current drops. As a response, current regulator ramps up the stator voltage reference further into the nonlinear region until the stator current resumes to a steady convergence to its reference. This can be seen in increased $V_{qs}^e$ and overmodulated duty cycles $d_{ABC}$. Another inevitalbe consequence of overmodulation is the emergence of non-triplen low-frequency harmonics, which distort the shapes of otherwise symmetric small wave ripples in qd-current, voltage and torque waveforms.

After the torque step the current and, consequently, voltage demand drops, bringing the VSI back to relative vicinity of the linear modulation region. Still some minor overmodulation can be seen in duty cycles and ripple shapes.

Overall, the simulation proved the regulator capability to operate in overmodulation conditions near the linear region, although the high-performance current control is deteriorated by delays and overshoots, as also stated in \cite{Hava}. From \cite{Hava} it is also known that one should not generalize the mentioned controllability conclusion in overmodulation conditions, since the regulator performance depends on numerous factors, such as the selected modulation technique, control features (anti-windup), DC voltage profile and load characteristics. Furthermore, overmodulation operation is not desired (and sometimes even prohibited \cite{Hava}) no only because of potential instability, but also due to increased low-frequency harmonic content in the IM and associated losses, torque ripple and so on \cite{Mahlfeld}.

\section{Conclusion and Outlook} \label{Sec:Conclusion}
\subsection{Conclusion}
Within the context of high-performance Indirect Field Oriented Control (IFOC), this work addresses the ramp-comparison current regulation in Pulse Width Modulated (PWM) Voltage Source Inverter-fed (VSI) Induction Machines (IM). So far, the simplest controller-modulator realization was analyzed both theoretically and numerically to gain a firm understanding of the system performance. A standard frequency-domain analysis along with IM model was used to design the PI current regulator. Matlab/Simulink simulations proved the validity of controller design. Next, high-speed/low voltage limitations were observed in regulator performance. Theoretical derivations and simulations associate those with inherent VSI DC voltage limit and with a simple modulator realization that results in poor DC voltage utilization. Nevertheless, it was shown that the regulator still operates in aforementioned conditions (up to a certain point) at the price of prolonged transients, overshoots, increased losses and ripple.

\subsection{Outlook}
Having studied the simplest controller-modulator realization, and observed its numerous limitations, this section briefly lists the potential improvements to be made:

$\bullet$ Advanced modulation strategies (SVM, DPWM and their combinations) can improve the output fundamental voltage and current, shorten the system dynamic response time, and expand steady-state operating region \cite{Holmes}, \cite{Mahlfeld},\cite{Guo}.

$\bullet$ Anti-windup and model-based disturbance feedforward can improve current regulator performance in saturation conditions \cite{Novotny}, \cite{Hava}. Prevention of unbounded growth of error integral is expected to reduce the dynamic overshoot of the regulator. Disturbance feedforward can significantly reduce the regulator burden and improve reference tracking.

$\bullet$ Dual-sampling dual-update can be used to reduce the impact of digital control delay and half-bridge dead-times \cite{Hava}, \cite{Yu}.

$\bullet$ Nonlinear effects of time-varying IM parameters (e.g. magnetizing inductance saturation) can be added to the model, improving its accuracy \cite{Novotny}.

$\bullet$ As it was shown in \eqref{eq:VqdsCondition}, IM speed is ultimately limited by the DC voltage even when advanced modulation and control techniques are applied. Assuming a constant DC voltage source feeding the VSI, a standard solution would be to add a DC-DC boost converter as a front end.

To preserve stable operation at high-speeds, not only the DC voltage, but also the cascaded controller bandwidths have to increase, leading to elevated switching frequency. This, in turn, potentially increases the switching losses and heatsink volume.

To resolve this trade off between high-speed operation and switching loss, \cite{Kolar} introduces the so-called "synergetic" control. The idea behind is simple: DC-DC boost converter could control the DC voltage to behave like a maximum line-to-line voltage, requested from AC side (similar to three-phase diode rectifier output). Such DC voltage shaping makes the switching of 2 out of 3 phases unnecessary, leaving only 1 VSI phase operating at any instant in time. In space vector notation, the VSI always operates at the edges of the hexagon, however, the average trajectory is circular (not hexagon) because the hexagon itself is "breathing" (pulsating, growing/shrinking).

As a future work, one could evaluate the impact of "synergetic" control on the IM voltage/current frequency content, losses etc. and compare it to the results of other modulation techniques.

\section*{Acknowledgment}

The author thanks Prof. Dr. Tom Jahns and Mr. Pablo Castro Palavicino for their valuable insights and comments both during lectures and discussion sessions.
\balance



\section*{Supplementary material}

# Appendix

# Main Code

```matlab
clear all;
close all;
clc;
%% Rated Values (rms)
VsR  = 460;             % Rated rms line-to-line voltage
PmR  = 1.4873e+04;      % Rated power (20 hp)
fsR  = 60;              % Rated fundamental frequency
fcar = fsR*100;         % PWM Carrier frequency
WsR  = 2*pi*fsR;        % Rated angular frequency
P    = 4;               % Number of poles
sR   = 0.03135;         % Rated slip
IsR  = 32.9660;         % Rated phase current amplitude
thr  = acos(0.861);     % Rated current phasor angle
WmR  = WsR*(1-sR)*(2/P); % Rated mechanical rotor angular speed
WrR  = WmR*(P/2);       % Rated electrical rotor angular speed
WslR = WsR-WrR;         % Rated slip angular speed
TmR  = PmR/WmR;         % Rated mechanical torque
%% Base Values
Vb   = 376;
Ib   = 26.5;
Zb   = 14.20;
Tb   = 79.16;
Pb   = 3*Vb*Ib/2;
Wb   = WsR;
Wmb  = WsR*(2/P);
%% Mechanical Parameters
M    = 01.400;
J    = Pb/(Wmb^2)*M;
b    = 00.000;
%% Electric Parameters (in Ohms and Henries)
rs   = 00.35500;
rr   = 00.35500;
Lm   = 34.100/Wb;
Lls  = 01.420/Wb;
Llr  = 01.420/Wb;
Ls   = Lls + Lm;
Lr   = Llr + Lm;
D    = Ls*Lr - Lm^2;
sig  = 1 - (Lm^2)/(Lr*Ls);
LM   = 1/(1/Lm + 1/Lls + 1/Llr);
%% Mechanical Parameters in pu
Jpu  = J*Wmb^2/Pb;
bpu  = b*Wmb^2/Pb;
%% Electric Parameters in pu
rs_pu  = rs/Zb;
rr_pu  = rr/Zb;
Xm_pu  = Wb*Lm/Zb;
Xls_pu = Wb*Lls/Zb;
Xlr_pu = Wb*Llr/Zb;
Xs_pu  = Xls_pu + Xm_pu;
Xr_pu  = Xlr_pu + Xm_pu;
D_pu   = Xs_pu*Xr_pu - Xm_pu^2;
XM_pu  = 1/(1/Xm_pu + 1/Xls_pu + 1/Xlr_pu);
```

```matlab
%% Initial Conditions in pu
% The inputs of the induction machine (working as motor) are going to be
% the amplitude and frequency of the stator voltage and the load torque
% applied in the shaft, with this the slip must be calculated and then the
% currents, speed and flux linkages.
%% Simulation Conditions in pu
RF    = 3; % Choose reference frame
% RF = 1 , stator reference frame
% RF = 2 , rotor reference frame
% RF = 3 , synchronous reference frame
tf    = 6/fsR; % Simulation time
T_Vs  = 6/fsR;  % Time to change stator voltage from Vs_o to Vs_oo
T_Tl  = 3/fsR;  % Time to change load torque from Tl_o to Tl_oo
T_Iq  = 3/fsR;  % Time to change q-axis stator current from Iqs_o to Iqs_oo
T_dc  = 7/fsR;  % Time to change Vdc from Vdc_o to Vdc_oo
ang_o = 0.224762;
Vs_oA = 1.00*VsR*sqrt(2/3)/Vb;
Vs_o  = Vs_oA*(cos(ang_o) + 1j*sin(ang_o));
Vs_oo = Vs_oA*(cos(ang_o) + 1j*sin(ang_o));
Ws_o  = 1.00*WsR;
Tl_o  = 1.0011*TmR/Tb;
Tl_oo = 0.5*Tl_o;
Vdc_o = 1.7*Vs_oA;
Vdc_oo= 1.7*Vs_oA;
%% ECE 411 method for steady state slip calculations in pu
% s    = linspace(0,0.1,1e7);
% Wr   = Ws_o*(1-s);
% Wm   = Wr*(2/P);
% Zs   = (rs    + 1j*Ws_o*Lls)/Zb;
% Zr   = (rr./s + 1j*Ws_o*Llr)/Zb;
% Zm   = (        1j*Ws_o*Lm)/Zb;
% Ztot = Zs + Zr.*Zm./(Zr + Zm);
% Is   = Vs_o./Ztot;
% Ir   =-Is.*Zm./(Zm + Zr);
% Pe   = (3/2)*(rr/Zb)*abs(Ir).^2.*(1-s)./s;
% Te   = Pe./(Wm/Wmb);
% % find point where Tl=Te
% a   = 0.0;
% k   = 1.0;
% tol = 0.0001;
% while (a==0)
%     if abs(Te(k)-Tl_o)<tol
%         N=k;
%         a=1;
%     end
%     k=k+1;
% end
%% HW6 method for steady state calculations in pu (contrast two methods!!!)
N    = 1;
s    = sR;
Wr   = Ws_o*(1-s);
Wm   = Wr*(2/P);
Zss  = rs_pu + 1i*(Xs_pu*Ws_o/Wb);
Zrr  = rr_pu + 1i*s*(Xr_pu*Ws_o/Wb);
Zsr  = 1i*Xm_pu*Ws_o/Wb;
Zrs  = 1i*s*Xm_pu*Ws_o/Wb;
Is   = Zrr*Vs_o./(Zss*Zrr - Zsr*Zrs);
```

```matlab
    Ir      =-Zrs*Vs_o./(Zss*Zrr - Zsr*Zrs);
    Te      = (3/2)*(P/2)*Lm*imag(Is.*conj(Ir))*(Ib^2)/Tb;
    Pe      = 1.5*real(Vs_o*Is)*Vb*Ib/Pb;
%% Initial Values in pu
Iqs_o       = real(Is(N));
Ids_o       =-imag(Is(N));
Iqr_o       = real(Ir(N));
Idr_o       =-imag(Ir(N));
Fqs_o       = (Ls*Iqs_o + Lm*Iqr_o)*Ib*Wb/Vb;
Fds_o       = (Ls*Ids_o + Lm*Idr_o)*Ib*Wb/Vb;
Fqr_o       = (Lm*Iqs_o + Lr*Iqr_o)*Ib*Wb/Vb;
Fdr_o       = (Lm*Ids_o + Lr*Idr_o)*Ib*Wb/Vb;
Wm_o        = Wm(N)/Wmb;
Wr_o        = Wr(N);
Thetar_o    = angle(Ir(N));
%% Current controller parameters
wc      = 2*pi*fcar/10;
Kp      = (Ib/Vb)*wc*sig*Ls;
Ki      = (Ib/Vb)*Kp*(rs + rr*(Lm/Lr)^2)/(sig*Ls);
%% Simulate system in pu
out = sim('VSI_IM_Model_pu');
%% Load Data From Simulink
t    = out.Results(:,01)*fsR;
Vas  = out.Results(:,02);
Vbs  = out.Results(:,03);
Vcs  = out.Results(:,04);
VqsC = out.Results(:,05);
VdsC = out.Results(:,06);
Ias  = out.Results(:,07);
Ibs  = out.Results(:,08);
Ics  = out.Results(:,09);
Iqs  = out.Results(:,10);
Ids  = out.Results(:,11);
Iar  = out.Results(:,12);
Ibr  = out.Results(:,13);
Icr  = out.Results(:,14);
dA   = out.Results(:,15);
dB   = out.Results(:,16);
dC   = out.Results(:,17);
Ws   = out.Results(:,18)/Wb;
Wr   = out.Results(:,19)/Wb;
Wm   = out.Results(:,20);
Te   = out.Results(:,21);
Tl   = out.Results(:,22);
IqsC = out.Results(:,23);
IdsC = out.Results(:,24);
% Fqs = out.Results(:,22);
% Fds = out.Results(:,23);
% Fqr = out.Results(:,24);
% Fdr = out.Results(:,25);
% Iin = out.Results(:,26);
% Idc = out.Results(:,27);
%% Plots in pu
% Change the ylim parameters for change the y-axis range so you can see
% the behavior of the signals better in case that your transient is very
% pronounced
% close all;
```

```matlab
% hFig01 = figure(01);set(hFig01,'Units','centimeters', 'Position', [00.50 00.50  10.00 24.0])
% hFig02 = figure(02);set(hFig02,'Units','centimeters', 'Position', [00.50 02.00  12.00 04.0])
% hFig03 = figure(03);set(hFig03,'Units','centimeters', 'Position', [13.50 09.00  12.00 04.0])
% hFig04 = figure(04);set(hFig04,'Units','centimeters', 'Position', [13.50 02.00  12.00 04.0])
% hFig05 = figure(05);set(hFig05,'Units','centimeters', 'Position', [26.50 09.00  12.00 04.0])
% hFig06 = figure(06);set(hFig06,'Units','centimeters', 'Position', [26.50 02.00  12.00 04.0])
% figure(01),plot(t ,Iqs,' b','linewidth',2),grid on,hold on,ylim([ 0.8 1.4]),xlim([10/fsR 12/fsR]),xlabel('Time [s]'),ylabel('i_q_s^e [pu]'    )
% figure(02),plot(t ,Ids,' r','linewidth',2),grid on,hold on,ylim([ 0.0 1.2]),xlim([10/fsR 12/fsR]),xlabel('Time [s]'),ylabel('i_d_s^e [pu]'    )
% figure(03),plot(t ,Iin,' k','linewidth',2),grid on,hold on,ylim([ 0.0 1.8]),xlim([10/fsR 12/fsR]),xlabel('Time [s]'),ylabel('I_i    [pu]'    )
% figure(04),plot(t ,Idc,' k','linewidth',2),grid on,hold on,ylim([ 0.8 1.2]),xlim([10/fsR 12/fsR]),xlabel('Time [s]'),ylabel('I_b    [pu]'    )
% figure(05),plot(t ,Te ,' g','linewidth',2),grid on,hold on,ylim([ 0.8 1.2]),xlim([10/fsR 12/fsR]),xlabel('Time [s]'),ylabel('T_e    [pu]'    )
% figure(06),plot(t ,Wr ,' b','linewidth',2),grid on,hold on,ylim([ 0.8 1.2]),xlim([10/fsR 12/fsR]),xlabel('Time [s]'),ylabel('\omega_r [pu]'  )
%% Subplot in pu
close all;
hFig01 = figure(01);set(hFig01,'Units','centimeters', 'Position', [00.50 02.50  10.00 12.0]);
figure(01),subplot(4,1,1)
plot(t ,Iqs ,'b','linewidth',1),grid on,hold on,xlabel('(a) Time ($T_s$)','interpreter','latex'),ylabel('$i_{qs}^e$ [pu]','interpreter','latex'),ylim([ 0.50 1.30])
plot(t ,IqsC,'y','linewidth',1),legend('$i_{qs}^e$','$i_{qs}^{e*}$','interpreter','latex')
figure(01),subplot(4,1,2)
plot(t ,Ids ,'b','linewidth',1),grid on,hold on,xlabel('(b) Time ($T_s$)','interpreter','latex'),ylabel('$i_{ds}^e$ [pu]','interpreter','latex'),ylim([ 0.20 0.50])
plot(t ,IdsC,'y','linewidth',1),legend('$i_{ds}^e$','$i_{ds}^{e*}$','interpreter','latex')
figure(01),subplot(4,1,3)
plot(t ,VqsC,'b','linewidth',1),grid on,hold on,xlabel('(c) Time ($T_s$)','interpreter','latex'),ylabel('$V_{qds}^e$ [pu]','interpreter','latex'),ylim([-0.50 1.40])
plot(t ,VdsC,'r','linewidth',1),legend('$V_{qs}^{e*}$','$V_{ds}^{e*}$','interpreter','latex')
figure(01),subplot(4,1,4)
plot(t ,dA  ,'b','linewidth',1),grid on,hold on,xlabel('(d) Time ($T_s$)','interpreter','latex'),ylabel('$d_{ABC}$','interpreter','latex'),ylim([-0.40 1.40])
plot(t ,dB  ,'g','linewidth',1)
plot(t ,dC  ,'r','linewidth',1),legend('$d_A$','$d_B$','$d_C$','interpreter','latex')
```

```matlab
hFig02 = figure(02);set(hFig02,'Units','centimeters', 'Position', [10.50 02.50  10.00 12.0]);
figure(02),subplot(4,1,1)
plot(t ,Vas ,'b','linewidth',1),grid on,hold on,xlabel('(e) Time ($T_s$)','interpreter','latex'),ylabel('$V_{as}$ [pu]','interpreter','latex'),legend('$V_{as}$','interpreter','latex'),ylim([-2.00 2.00])
figure(02),subplot(4,1,2)
plot(t ,Ias ,'b','linewidth',1),grid on,hold on,xlabel('(f) Time ($T_s$)','interpreter','latex'),ylabel('$i_{abcs}$ [pu]','interpreter','latex'),ylim([-1.50 1.50])
plot(t ,Ibs ,'g','linewidth',1)
plot(t ,Ics ,'r','linewidth',1),legend('$i_{as}$','$i_{bs}$','$i_{cs}$','interpreter','latex')
figure(02),subplot(4,1,3)
plot(t ,Iar ,'b','linewidth',1),grid on,hold on,xlabel('(g) Time ($T_s$)','interpreter','latex'),ylabel('$i_{abcr}$ [pu]','interpreter','latex'),ylim([-1.50 1.50])
plot(t ,Ibr ,'g','linewidth',1)
plot(t ,Icr ,'r','linewidth',1),legend('$i_{ar}$','$i_{br}$','$i_{cr}$','interpreter','latex')
figure(02),subplot(4,1,4)
plot(t ,Te  ,'b','linewidth',1),grid on,hold on,xlabel('(h) Time ($T_s$)','interpreter','latex'),ylabel('$T_{e}, T_l$ [pu]','interpreter','latex'),ylim([ 0.40 1.20])
plot(t ,Tl  ,'r','linewidth',1),legend('$T_e$','$T_l$','interpreter','latex')
%% Bode plots
% close all;
% clear all;
% clc;
% wc    = 2*pi*60*10;
% Tol   = tf(wc,[1 0]);
% Tcl   = 1/(1 + 1/Tol);
% hFig01 = figure(01);set(hFig01,'Units','centimeters', 'Position', [00.50 09.00  10.00 06.00])
% figure(01),hold on,bodeplot(Tol),h=bodeplot(Tcl),grid on,legend('$T\_{ol}(s)$','$T\_{cl}(s)$','interpreter','latex');
% options = getoptions(h);
% options.FreqUnits = 'Hz';
% options.Title.String = ' ';
% options.XLabel.Interpreter = 'latex';
% options.XLabel.FontSize = 9;
% options.YLabel.Interpreter = 'latex';
% options.YLabel.FontSize = 9;
% options.TickLabel.FontSize = 9;
% setoptions(h,options);
```

## Inverter Switching Function

```matlab
function [Va,Vb,Vc,Iin] = Inverter(Vdc,Ia,Ib,Ic,Sw1,Sw3,Sw5)
Iin = Sw1*Ia + Sw3*Ib + Sw5*Ic;
Va  = 0;
Vb  = 0;
Vc  = 0;
if Sw1==0 && Sw3==0 && Sw5==0, Va= (0/3)*Vdc; Vb= (0/3)*Vdc; Vc= (0/3)*Vdc;end
if Sw1==1 && Sw3==0 && Sw5==0, Va= (2/3)*Vdc; Vb=-(1/3)*Vdc; Vc=-(1/3)*Vdc;end
if Sw1==1 && Sw3==1 && Sw5==0, Va= (1/3)*Vdc; Vb= (1/3)*Vdc; Vc=-(2/3)*Vdc;end
if Sw1==0 && Sw3==1 && Sw5==0, Va=-(1/3)*Vdc; Vb= (2/3)*Vdc; Vc=-(1/3)*Vdc;end
if Sw1==0 && Sw3==1 && Sw5==1, Va=-(2/3)*Vdc; Vb= (1/3)*Vdc; Vc= (1/3)*Vdc;end
if Sw1==0 && Sw3==0 && Sw5==1, Va=-(1/3)*Vdc; Vb=-(1/3)*Vdc; Vc= (2/3)*Vdc;end
if Sw1==1 && Sw3==0 && Sw5==1, Va= (1/3)*Vdc; Vb=-(2/3)*Vdc; Vc= (1/3)*Vdc;end
if Sw1==1 && Sw3==1 && Sw5==1, Va= (0/3)*Vdc; Vb= (0/3)*Vdc; Vc= (0/3)*Vdc;end
end
```

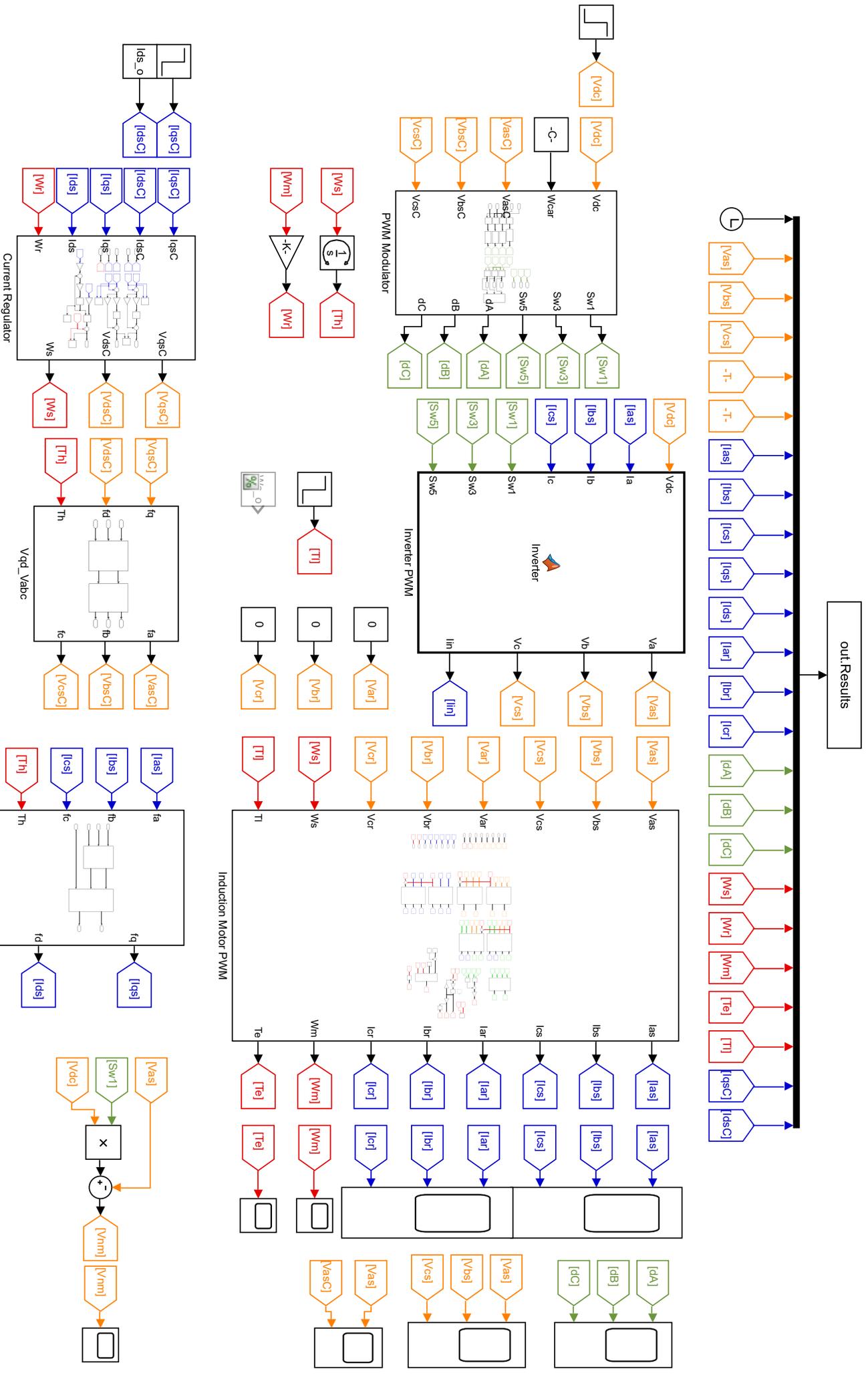

Voltage Source Inverter - Induction Machine Model in [pu]

# Current Regulator

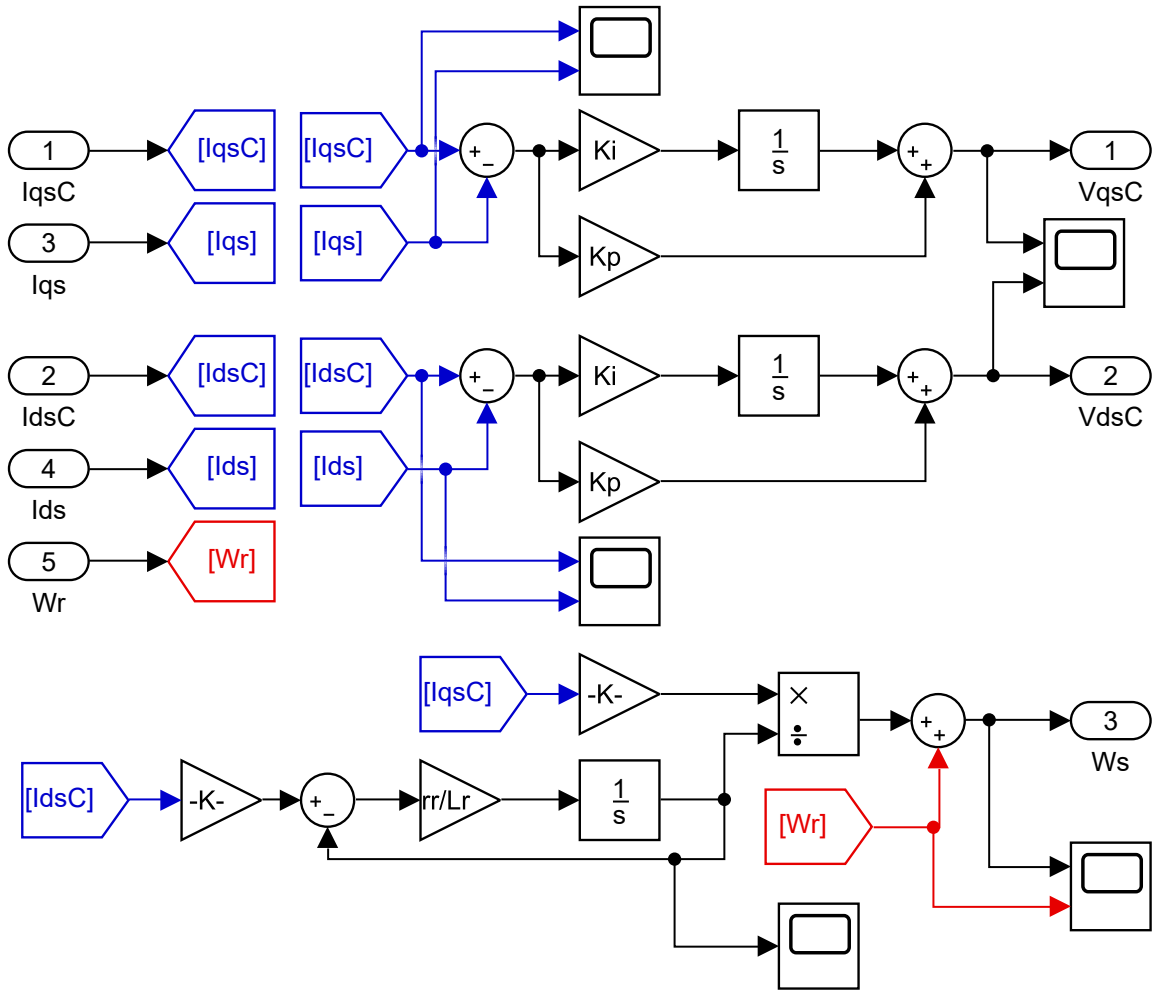

# Synchronous Frequency Estimator (IFOC)

# PWM Modulator

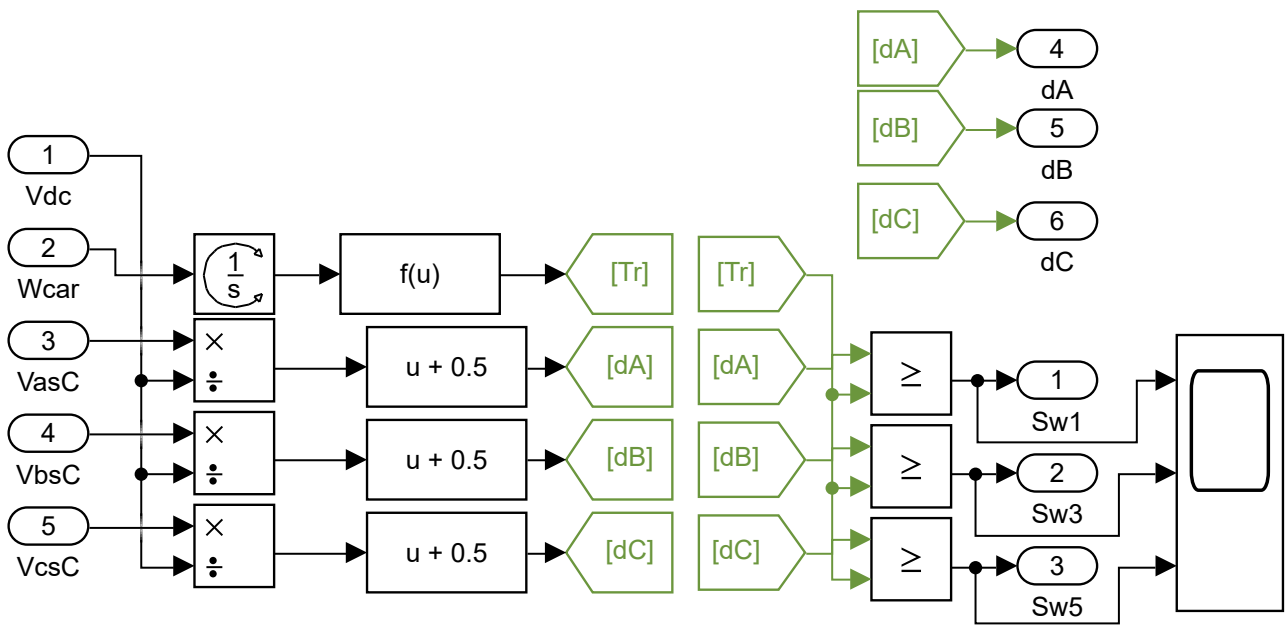

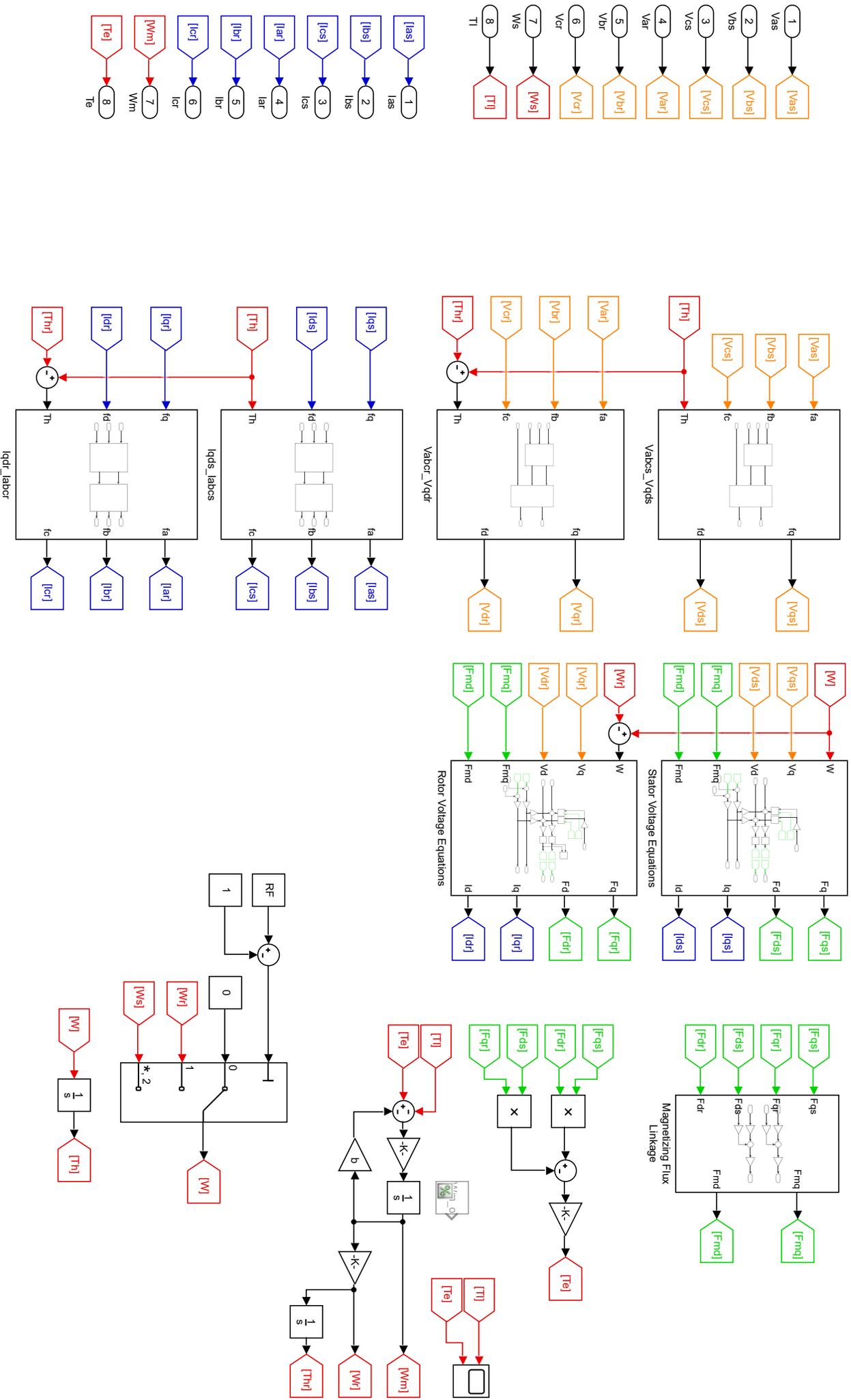

Induction Machine Model in [pu]




\begin{thebibliography}{00}
	
\bibitem{Abad}  G. Abad, J. Lopez, M. A. Rodr ´ ´ıguez, L. Marroyo, and G. Iwanski, Doubly Fed Induction Machine: Modeling and Control for Wind Energy Generation. New York, NY, USA: Wiley, 2011.
\bibitem{Pellegrino} G. Pellegrino, A. Vagati, B. Boazzo and P. Guglielmi, "Comparison of Induction and PM Synchronous Motor Drives for EV Application Including Design Examples," IEEE Trans. Industry Applications, vol. 48, pp. 2322-2332, Nov.-Dec. 2012.
\bibitem{Novotny} D. Novotny and T. Lipo, \textit{Vector Control and Dynamics of AC Drives}, ser. Monographs in electrical and electronic engineering. Clarendon Press, 1996, no. v. 1.
\bibitem{Holmes} D. G. Holmes, T. A. Lipo, "Pulse Width Modulation For Power
Converters – Principles and Practice," IEEE press, 2003. 
\bibitem{Hava} A. M. Hava, “Carrier based PWM-VSI overmodulation strategies,”
Ph.D. dissertation, Univ. Wisconsin, Madison, 1997.
\bibitem{Rowan} T. Rowan, "Analysis of /naturally Sampled Current Regulated Pulse-width Modulated Inverters", Ph.D. Thesis, University of Wisconsin, 1985.
\bibitem{Shin} Eun-Chul Shin, Tae-Sik Park, Won-Hyun Oh and Ji-Yoon Yoo, "A design method of PI controller for an induction motor with parameter variation," \textit{IECON'03. 29th Annual Conference of the IEEE Industrial Electronics Society (IEEE Cat. No.03CH37468)}, Roanoke, VA, USA, 2003, pp. 408-413 vol.1.
\bibitem{Ogata} Katsujiko Ogata, \textit{Modern Control Engirierring}, 2nd, Prentice-Hall, 1990, 595-605.
\bibitem{Mahlfeld} H. Mahlfeld, T. Schuhmann, R. Döbler and B. Cebulski, "Impact of overmodulation methods on inverter and machine losses in voltage-fed induction motor drives," 2016 XXII International Conference on Electrical Machines (ICEM), Lausanne, 2016, pp. 1064-1070.
\bibitem{Guo} X. Guo, M. He and Y. Yang, "Over Modulation Strategy of Power Converters: A Review," in IEEE Access, vol. 6, pp. 69528-69544, 2018.
\bibitem{Yu} Y. Yu, Q. Dong, Z. Dong, Y. Shao, B. Wang and D. Xu, "Optimized Current Regulator Design for Induction Motor Dynamic Performance Improvement in Over-Modulation Region," 2018 21st International Conference on Electrical Machines and Systems (ICEMS), Jeju, 2018, pp. 1187-1192.
\bibitem{Kolar} J. W. Kolar, M. Guacci, M. Antivachis, D. Bortis, Next Generation 3-Phase Variable-Speed Drive PWM Inverter Concepts, Keynote Presentation at the 20th International Symposium POWER ELECTRONICS (Ee2019), Novi Sad, Serbia, October 23-26, 2019.
\end{thebibliography}
\end{document}